\documentclass{ws-p8-50x6-00}
\usepackage{psfig}
\begin{document}

\title{High redshift CO line emission: perspectives}
\author{F. Combes}

\address{DEMIRM, Observatoire de Paris, \\
61 Av. de l'Observatoire, F-75 014, Paris, France\\
E-mail: francoise.combes@obspm.fr}

\maketitle

\abstracts{
Although about a dozen high redshift ($z$ larger than 2) starburst galaxies
have been recently detected in the CO lines, spectroscopic detections
of molecular gas of very young galaxies are still very difficult.
The CO lines are usually optically thick, which limits greatly the increase
of emission with redshift, as observed for the dust continuum. However
the future instruments (LMT, ALMA, etc..) will allow large progress in this
domain, and perspectives are discussed. Computations are based on
a simple extrapolation of what is known of starbursting galaxies
at lower redshift.}

\section{Introduction}

One of the breakthrough due to recent progress on faint 
galaxies has been the inventory of the amount of star
formation at every epoch (e.g. Madau et al. 1996).
 The comoving star formation rate appears to increase like
$(1+z)^4$ from $z = 0$ to $z = 1$, and then decreases again
to the same present value down to $z = 5$. But this relies
on the optical studies, i.e. on the UV-determined star
forming rates in the rest-frame. If early starbursts are dusty,
this decrease could be changed into a plateau
(Guiderdoni et al. 1997; Blain et al. 1999).
 To tackle this problem, independent information 
must be combined in a coherent picture, such as that coming
from the far-infrared and millimeter domains: 
the cosmic IR and submm background 
radiation discovered by COBE (e.g. Puget et al. 1996;
Hauser et al. 1998) yields an insight on the global
past star-formation of the Universe, and the sources discovered
at high redshift in the millimeter continuum and lines 
yield information on the structure of the past starbursts
(Smail et al. 1997; Hughes et al. 1998; Barger et al. 1999; 
Guilloteau et al. 1999).

This review focus on the molecular content of galaxies,
that we can deduce from CO lines. Although the
submillimeter continuum is more easy to detect at high
redshift, it cannot without ambiguity 
trace the evolution of star formation as a function
of redshift, because of identification problems, unknown
redshifts, and unknown contribution of AGN.
After summarising the present state of knowledge 
concerning CO emission lines,
perspectives are drawn concerning the future surveys
that will be conducted with the next generation of millimeter
instruments.

\section{CO Detections at High Redshift}  

 The detection of highly redshifted millimeter CO lines in the
hyperluminous object IRAS 10214+4724
($z=2.28$, Brown \& Vanden Bout 1992, Solomon et al. 1992), has opened
this new field, to explore the history of star formation, and its
efficiency. Although the first enormous derived H$_2$ mass
has now been revised (the source is considerably amplified by 
lensing), it is still surprising to find such huge amounts of CO
molecules, especially since the gas is expected to have lower 
metallicity at high z.  But theoretical calculations have shown that in a violent
starburst, the metallicity could reach solar values very quickly
(Elbaz et al. 1992).

Today, more than a dozen objects have been detected in CO
lines at high z: they are often gravitationally amplified,
either being multi-imaged by a strong lens, like 
the Cloverleaf quasar H 1413+117 at $z=2.558$ (Barvainis et al. 1994),
the lensed radiogalaxy MG0414+0534 at $z=2.639$ (Barvainis et al. 1998),
or the possibly magnified object
BR1202-0725 at $z=4.69$ (Ohta et al. 1996, Omont et al. 1996); or
they are more weakly amplified by a foreground galaxy cluster, like
the submillimeter-selected hyperluminous galaxies SMM02399-0136
at $z=2.808$ (Frayer et al. 1998), and SMM 02399-0134at 1.062
(Kneib et al. 2000). Often several high-J CO lines are detected,
revealing the high temperature of the gas (typical of a starburst
at $\sim$ 60K). Higher temperatures are rare, as in
the magnified BAL quasar APM08279+5255,
at $z=3.911$, where the gas temperature derived from the CO lines is
$\sim$ 200K, maybe excited by the quasar (Downes et al. 1999).
Recently Scoville et al. (1997b) reported the detection of the first
non-lensed object at $z=2.394$, the weak radio galaxy 53W002,
and Guilloteau et al. (1997) the radio-quiet quasar BRI 1335-0417, 
at $z=4.407$, which has no direct indication of lensing.
If the non-amplification is confirmed, these objects
 would contain the largest molecular contents known
($8-10 \cdot 10^{10}$ M$_\odot$ with a standard CO/H$_2$
conversion ratio, and even more if the metallicity is low).
The derived molecular masses are so high that H$_2$ would constitute
between 30 to 80\% of the total dynamical mass (cf 4C 60.07, 
Papadopoulos et al. 2000), if the standard CO/H$_2$ conversion ratio was 
adopted. The application of this conversion ratio is however doubtful, and 
it is possible that the involved H$_2$ masses are 3-4 times lower (Solomon
et al. 1997).

To search for primeval galaxies, already Elston et al. (1988)
had identified extremely red objects that are conspicuous only in the
near-infrared (R - K $>$ 5). 
Maybe 10\% of the submm sources could be EROs
(Smail et al. 1999).  A proto-typical ERO at $z$=1.44 
(Dey et al. 1999) has been detected in submm continuum 
(Cimatti et al. 1998), and in CO lines (Andreani et al. 2000).

\begin{figure}
\centering
\psfig{figure=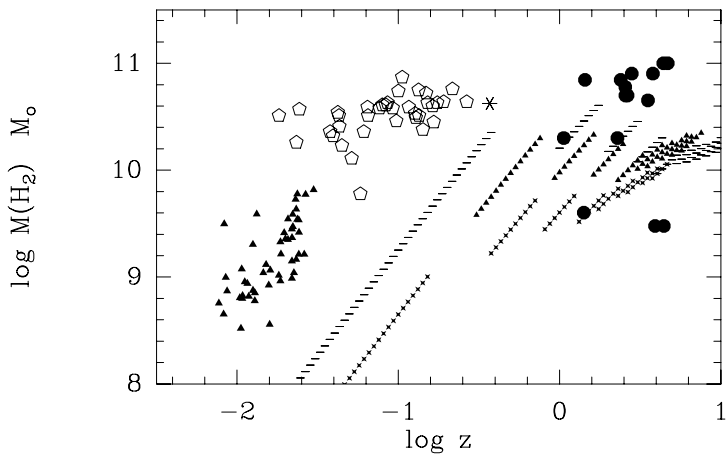,bbllx=2cm,bblly=1cm,bburx=12cm,bbury=16cm,angle=-90,width=10cm}
\caption{H$_2$ masses for the CO-detected objects at high redshift 
(full dots), compared to the ultra-luminous-IR sample of Solomon et al. (1997,
open pentagons), to the Coma supercluster sample from Casoli et
al (1996, filled triangles), and to the quasar
3c48, marked as a star (Scoville et al. 1993, Wink et al. 1997). The lines
delineated by various symbols
indicate the 1$\sigma$ detection limit at the IRAM-30m
telescope of  S(CO) = 1.0 Jy km/s with the 3mm receiver (hyphens),
2.0  Jy km/s with the 2mm (triangles) and 1.3mm (crosses) receivers,
rms that we can reach in 6h integration.
Note the absence of detected objects between  0.36 and 1 in redshift.
The points at high $z$ can be detected well below the sensitivity
 limit, since they are gravitationally amplified. }
\label{mh2z}
\end{figure}

The CO line detections at high $z$ up to now are summarized in 
Table \ref{COdata}, and 
the molecular masses as a function of redshift are
displayed in Fig. \ref{mh2z}. It is clear from this figure that
only the very strongest objects are detected, and in 
particular the gravitationally amplified ones.
 but this will rapidly change with
the new millimeter instruments planned over the world
(the Green-Bank-100m of NRAO, the LMT-50m of UMass-INAOE,
the ALMA (Europe/USA) and the
LMSA (Japan) interferometers). It is therefore interesting to
predict with simple models the detection capabilities, as a function
of redshift, metallicity or physical conditions in the high-z objects.

\vspace*{-0.3cm}
\section{Perspectives with Future mm Instruments}

If so many sources have been detected now in the 
submillimeter domain at high redshifts,
 it is because around 1mm, starbursts have a larger
apparent flux at $z=2$ or 3  than $z=1$.
The line emission does not have such a negative
K-correction, since in the low frequency domain, the
flux of the successive lines increases roughly as $\nu^2$
(optically thick domain), instead of  $\nu^4$ for the 
continuum. 
Nevertheless the line emission is essential to study
the nature of the object (the AGN-starburst connection
for instance), and deduce more physics (kinematics,
abundances, excitation, etc..). Given the gas and dust 
temperatures, the maximum flux is always reached at
much lower frequencies than in the continuum, since the lines
always reflect the energy difference between two levels; 
this is an advantage, given the largest atmospheric 
opacity at high frequencies.

\begin{table}[h]
\caption[ ]{CO data for high redshift objects}
\begin{flushleft}
\begin{tabular}{lllclcl}  \hline
Source    & $z$   &  CO  & S  & $\Delta$V& MH$_2$   & Ref  \\
                &       &line  & mJy  & km/s & 10$^{10}$ M$_\odot$    &           \\
\hline
SMM 02399-0134&1.062&2-1 & 3 & 500    &   2$^*$     & 1  \\
0957+561        & 1.414 & 2-1  & 3  &  440 &  0.4$^*$ & 2  \\
HR10               & 1.439 & 2-1  & 4  &  400 &    7          & 3  \\
F10214+4724  & 2.285 & 3-2  & 18 & 230  & 2$^*$      &  4   \\
53W002          & 2.394 & 3-2  &  3 & 540  & 7              &  5   \\
H 1413+117    & 2.558 & 3-2  & 23 & 330  & 2-6 $^*$  &  6   \\
SMM 14011+0252&2.565& 3-2  & 13 & 200  & 5$^*$    &  7   \\
MG 0414+0534& 2.639 & 3-2  &  4 & 580  & 5$^*$       &  8   \\
SMM 02399-0136&2.808& 3-2  &  4 & 710  & 8$^*$      &  9   \\
6C1909+722     &3.532& 4-3  &  2 & 530  & 4.5           &  10   \\
4C60.07           &3.791& 4-3  &  1.7 & 1000  & 8           &  10   \\
APM 08279+5255&3.911& 4-3  &  6 & 400  & 0.3$^*$    &  11   \\
BR 1335-0414& 4.407 & 5-4  &  7 & 420  & 10              &  12   \\
BR 0952-0115& 4.434 & 5-4  &  4 & 230  & 0.3$^*$      &  13   \\
BR 1202-0725& 4.690 & 5-4  &  8 & 320  & 10              &  14   \\
\hline 
\end{tabular}
\end{flushleft}
$^*$ corrected for magnification, when estimated\\
Masses have been rescaled to $H_0$ = 75km/s/Mpc. When multiple images
are resolved, the flux corresponds to their sum\\
(1) Kneib et al. (2000); (2) Planesas et al. (1999)
(3) Andreani et al. (2000); (4) Solomon et al. (1992), 
Downes et al. (1995); (5) Scoville et al. (1997b); 
(6) Barvainis et al. (1994); (7) Frayer et al. (1999);
(8) Barvainis et al. (1998); (9) Frayer et al. (1998); 
(10) Papadopoulos et al. (2000);
(11) Downes et al. (1999); (12) Guilloteau et al. (1997); 
(13) Guilloteau et al. (1999); (14) Omont et al. (1996)
\label{COdata}
\end{table}

\subsection{Predicted Line and Continuum Fluxes}

To model high-redshift starburst objects, let us
extrapolate the properties of more local ones:
the active region is generally confined to a compact nuclear
disk, sub-kpc in size (Scoville et al. 1997a),
Solomon et al. 1990, 97). The gas is much denser
here than in average over a normal galaxy, of the order
of 10$^4$ cm$^{-3}$, with clumps at least of  10$^6$ cm$^{-3}$
to explain the data on high density tracers (HCN, CS..).
To schematize, the ISM maybe modelled by two
density and temperature components, at 30 and 90K  
(cf. Combes et al. 1999). The total molecular mass
considered will be $6 \cdot 10^{10}$ M$_\odot$ and the 
average column density N(H$_2$) of 10$^{24}$ cm$^{-2}$,
typical of the Orion cloud center. 

Going towards high redshift ($z > 9$), the temperature of
the cosmic background T$_{\rm bg}$ becomes of the same order
as the interstellar dust temperature, and the excitation of
the gas by the background radiation competes with that
of gas collisions. It might then appear easier to detect the lines
(Silk \& Spaans 1997), but this is not the case
when every effect is taken into account. To have an idea
of the increase of the dust temperature with $z$, the simplest
assumption is to consider the same heating power due
to the starburst. At a stationary state, the dust must then radiate
the same energy in the far-infrared that it receives from
the stars, and this is proportional to the quantity
  $T_{\rm dust}^6 - T_{\rm bg}^6$, if the dust is optically thin, and
its opacity varies in $\nu^{\beta}$, with $\beta$ = 2. Keeping this 
quantity constant means that the energy re-radiated by the dust,
proportional to  $T_{\rm dust}^6$, is always equal to the energy it
received from the cosmic background,
proportional to  $T_{\rm bg}^6$, plus the constant energy flux
coming from the stars. Since $\beta$ can also be equal to 
1 or 1.5, or the dust be optically thick, we have also 
considered the possibility of keeping  $T_{\rm dust}^4 - T_{\rm bg}^4$
constant; this does not change fundamentally
the results.

\begin{figure}
\psfig{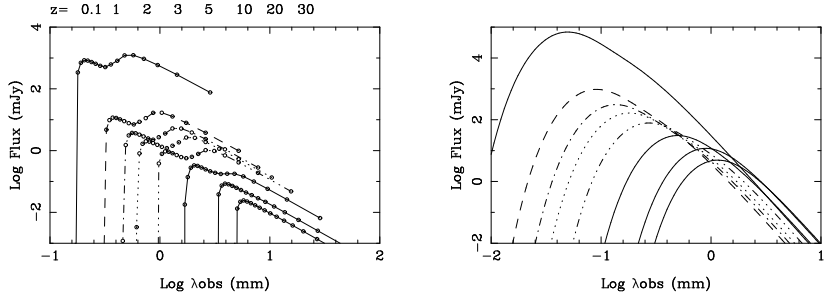}
\caption{ Expected flux for the two-component cloud model, for various
redshifts $z$ = 0.1, 1, 2, 3, 5, 10, 20, 30, and $q_0$ = 0.5.
{\bf Left} are the CO lines, materialised each by a circle
(they are joined by a line only to guide the eye). 
{\bf Right} is the continuum emission from dust. It has been assumed here 
that $T_{\rm dust}^6 - T_{\rm bg}^6$ is conserved
(from Combes et al. 1999).}
\label{cohz_39}
\end{figure}

Computing the populations of the CO rotational levels
with an LVG code, and in the case of the two component
models described earlier, the predictions of the line
and continuum intensities as a function of redshift and
frequencies are plotted in Fig. \ref{cohz_39}.

\subsection{Source Counts}

To predict the number of sources that will become available
with the future sensitivity, let us adopt a simple model of
starburst formation, in the frame of the hierarchical theory
of galaxy formation. The cosmology adopted here is
an Einstein-de Sitter model, $\Omega$ = 1, with no
cosmological constant, and $H_0$ = 75 km/s/Mpc, 
$q_0$ = 0.5.The number of mergers as a function 
of redshifts can be easily computed through the Press-Schechter
formalism (Press \& Schechter 1974), assuming 
self-similarity for the probability of dark halos merging,
and an efficiency of mergers in terms of star-formation
peaking  at $z \sim 2$ (i.e. Blain \& Longair 1993).
Also, the integration over all redshifts of the flux of all sources
should agree with the cosmic infrared background detected
by COBE. These contribute to reduce the 
number of free parameters of the modelling. To fit the source
counts, however, another parameter must be introduced 
which measures the rate of energy released in a merger
(or the life-time of the event): this rate must increase
strongly with redshift (cf. Blain et al. 1999).
Once the counts are made compatible with the submm
observations, the model indicates what must be
the contributions of the various redshift classes to the
present counts. It is interesting to
note that the intermediate redshifts dominate the
continuum source counts ($2 < z < 5$), if we allow the star formation
to begin before $z = 6$. At higher dust temperature, 
the counts are dominated by the highest redshifts 
($ z > 5$). 

\begin{figure}
\psfig{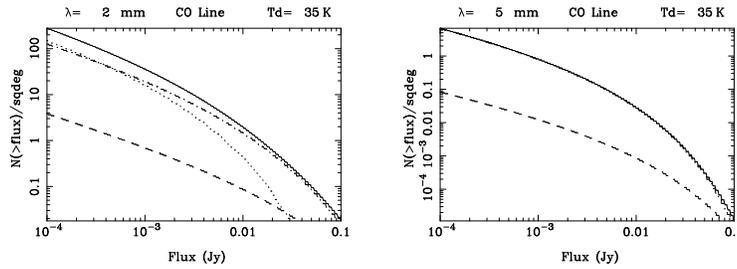}
\caption{ {\it Left} Source counts for the CO lines at an observed frequency 
of 2mm, assuming optically thick gas at T$_{ex}$ = 35K.
 The solid line is the total. Dash is the lowest
redshifts ($z < 2$); Dot-dash, intermediate ($2 < z < 5$);
Dots are the highest redshifts  ($z > 5$).
   {\it Right} Same for $\lambda$ = 5mm. }
\label{scoline}
\end{figure}

Once these fits are obtained for the continuum sources,
it is possible to derive also the counts for the CO line emission.
The spectral energy distribution is now obtained
with a comb-like function, representing the rotational ladder,
convolved with a Planck distribution of
temperature equal to the dust temperature, assuming the lines optically
thick. The frequency filling factor is then proportional to the 
rotational number, and therefore to the redshift, for a given observed 
frequency. The width of the lines have been assumed to be 300km/s. 
The derived source numbers are shown in Fig. \ref{scoline}.
 Note how they are dominated
by the high redshift sources. It is not useful to 
observe at $\lambda$ below 1mm for high-$z$ protogalaxies, but
instead to shift towards $\lambda$ = 1cm.

\section{Conclusion}

The search for CO lines in high redshift galaxies is only 
beginning, there will be a real breakthrough in the next decade,
when the sensitivity is increased by more than a factor 10
with the next generation millimeter instruments. 
CO lines are fundamental for our knowledge of star formation 
and efficiency at high z, to measure H$_2$ masses, recognize
AGN from starbursts, measure the dynamics and total masses
of galaxies.

\section*{Acknowledgements}
 I am very grateful to James Loventhal and David Hughes for the organisation of 
such an interesting meeting, and to the UMass for their financial help.

\end{document}